\documentclass{aa}
\usepackage{graphicx}
\usepackage{color}
\usepackage{natbib}
\usepackage{txfonts}

\newcommand{\eg}    {e.\,g.}
\newcommand{\ie}    {i.\,e.}
\newcommand{\sii}   {[S\,{\sc ii}]}

\newcommand{\fe}    {[Fe\,{\sc ii}]}
\newcommand{\hmol}  {H$_2~\upsilon=1$--0 S(1)}

\newcommand{\kms}   {km~s$^{-1}$}

\begin{document}

\title{
HH~223: a parsec-scale H$_2$ outflow in the star-forming region L723\thanks{
Based on observations made with LIRIS at the  4.2~m Williams Herschel Telescope
operated at the Observatorio del Roque de los Muchachos of the Instituto de
Astrof\'{\i}sica de Canarias. Fits files of Table 1 are available at
CDS via anonymous ftp to cdsarc.u-strasbg.fr (130.79.128.5)}
}

\author{R. L\'opez \inst{1}
\and J. A. Acosta-Pulido \inst{2,3}
\and G. G\'omez \inst{2,4} 
\and R. Estalella \inst{1} 
\and C. Carrasco-Gonz\'alez \inst{5}
}

\offprints{R. L\'opez}

\institute{
Departament d'Astronomia i Meteorologia (IEEC-UB), Institut de Ci\`encias
del Cosmos, Universitat de Barcelona, Mart\'{\i} i Franqu\`es 1, 
E-08028 Barcelona, Spain; email:rosario@am.ub.es,
robert.estalella@am.ub.es 
\and 
Instituto de Astrof\'{\i}sica de Canarias, E38200
La Laguna, Tenerife, Spain; email: jap@iac.es
\and
Departamento de  Astrof\'{\i}sica, Universidad de La Laguna, E38205, La Laguna,
Tenerife.
\and
GTC Office; email: gabriel.gomez@gtc.iac.es
\and
Instituto Astrof\'{\i}sica Andaluc\'{\i}a, CSIC, Camino
Bajo de Hu\'etor 50, E-18008 Granada, Spain; email: charly@iaa.es
}
\date{\today}

\titlerunning{NIR imaging of HH~223}

\abstract
{The dark cloud Lynds 723 (L723) is a low-mass star-forming region where one of
the few known cases of a quadrupolar CO outflow has been reported. Two recent
works have found that the radio continuum source VLA~2, towards the centre of
the CO outflow, is actually a multiple system of young stellar objects (YSOs).
Several line-emission nebulae that lie projected on the east-west CO outflow were
detected in narrow-band H$\alpha$ and \sii\ images. The spectra of the knots are
characteristic of shock-excited gas (Herbig-Haro spectra), with supersonic
blueshifted velocities, which suggests an optical outflow also powered by
the VLA~2 YSO system of L723.}
{Our aim is to study L723 in the near-infrared and look for line-emission nebulae
associated with the optical and CO outflows.}
{We imaged a field of $\sim~5~\arcmin \times 5~\arcmin$  centred on HH~223,
which includes the whole region of the quadrupolar CO outflow with
narrow-band  filters centred on the \fe\ 1.644~$\mu$m and H$_2$  2.122~$\mu$m
lines, together with  off-line $H_c$ and $K_c$ filters.  The \fe\
and H$_2$ line-emission structures were identified 
after extracting the continuum contribution, if any.
Their positions were determined from an accurate astrometry of the images.}
{The H$_2$ line-emission structures appear distributed over a region of $5\farcm5$ 
($\sim0.5$~pc for a distance of 300~pc)  at both sides of the  VLA~2 YSO
system, with an S-shape morphology,
and are projected onto the east-west CO outflow. Most
of them were resolved  in smaller knotty substructures. The  \fe\ emission only
appears associated with HH~223. An additional nebular emission from the 
continuum in
{\it H}$_c$ and {\it K}$_c$ appears associated with HH~223-K1, the structure
closest to the VLA~2 YSO system, and could be tracing the cavity walls.}
{We propose that the H$_2$ structures form part of a large-scale near-infrared
outflow, which is also associated with
the VLA~2 YSO system. The current data do not allow us to discern which of the
YSOs of VLA~2 is powering this large scale optical/near-infrared outflow.}


\maketitle

\section{Introduction}

\object{Lynds 723} (L723) is an isolate dark cloud located at a distance of
$300\pm150$ pc \citep{gol84} that shows evidence of low-mass star formation, 
and
is the site where one of the few known cases of a quadrupolar CO outflow \cite[two
separate pairs of red-blue lobes;][and references therein]{lee02}  has been
reported. The 3.6~cm radio continuum source \object{VLA~2} \citep{ang96},
towards the centre of the CO outflow, harbours the source that powers the
outflow.  VLA~2 is embedded in high-density gas traced by NH$_3$, which shows
evidence of gas heating and line broadening \citep{gir97}. Two recent works by
\citet{Car08} and \citet{Gir09} report that VLA~2 is a multiple system.
\citet{Car08} detect a system of four (VLA~2A, 2B, 2C and 2D) young stellar
objects (YSOs) and propose that the morphology of the CO outflow is actually the
result of the superposition of three independent pairs of CO lobes. They also
propose that one of the YSOs (\object{VLA~2A}) is powering the largest,
east-west pair of CO lobes and the system of emission-line optical nebulae,
which is
reminiscent of Herbig-Haro (HH) objects, first reported by \citet{vrb86}.
\citet{Gir09} detect at 1.35~mm emission from dust, resolved into two components,
\object{SMA1} and \object{SMA2}, which have very similar physical properties. SMA2
seems to be in a more evolved stage and is harbouring a multiple low-mass
protostellar system  (VLA~2A, 2B and 2C). They also report emission from the SiO
5--4 line towards the SMA sources, this emission shows an elongated morphology
that
follows the northwest-southeast direction pointed by  the larger CO outflow
lobes, which traces a region of interaction between the dense envelope and the
outflow.

In the optical wavelength range, deep H$\alpha$ and \sii\ narrow-band images of
the L723 field were obtained  by \citet{lop06}.  From these images,  the knotty
structure of the \object{HH~223} ``linear emission feature'' (hereafter,
we will refer to it as  HH~223)  detected by \citet{vrb86} was resolved  into
several knots, HH~223-A to -F, and the line-emission nature of other nebulae in
the field  was established. Long-slit spectroscopy  covering the spectral range
5800--8300 \AA\  was made by \citet{Lop09} through HH~223. The  spectra obtained
are  characteristic of shock-excited gas (HH spectra), both  for the knots and
for the low-brightness nebula surrounding the knots.  The radial velocities
derived for the knots are supersonic (blueshifted velocities, ranging from $-60$
to $-130$~\kms). The velocities derived for the low-brightness nebula are
compatible with the ambient gas velocity.

In the near-infrared range, \citet{pal99} imaged  a field of $\sim 2$~\arcmin\
centred on VLA~2 in the {\it K} band and detected H$_2$ emission from several
nebulae located at both sides of VLA~2. However, the field imaged only partially
covers L723: it does not include the complete region encompassed by the CO
outflow and, in particular, only the western side of HH~223  was mapped. 
Furthermore, since the continuum was not subtracted from their narrow-band H$_2$
image, it is impossible to know whether there is continuum contribution to the
emission in the nebula close to VLA~2. Continuum emission around this position
is expected if the nebula is tracing a cavity opened by the molecular outflow.

We imaged the L723 field with narrow-band filters centred on the \fe\
$\lambda$~1.644~$\mu$m and  \hmol\ 
($\lambda$~2.122~$\mu$m) lines together with  the broad-band {\it H} and the
off-line narrow-band {\it H}$_c$ and {\it K}$_c$ filters. Our aims were to get
a more complete picture of the HH~223 outflow in the near-infrared wavelength
range, to  search for counterparts of the optical nebulae, and to establish the
nature of the near-infrared emission coming from the nebulae of the L723 field.
The main results are presented in this paper.

\begin{table*}
\caption[ ]{Log of the observations}
\label{tlog}
\begin{tabular}{llcrrc}  
\hline\hline
           &$\lambda_c$/$\Delta\lambda$&Date&t$_{DIT}$&t$_{exp}$&seeing\\
Filter     &($\mu$m)/(\AA)  &2006/07/ &  (s) & (s) &(arcsec) \\        
\hline 
\fe 	      &  1.644/280 & 15 & 100 & 2000 & 1.05 \\
	          &            & 18 & 100 & 600  & 0.78 \\
	          &            & 20 &  60 & 1800 & 0.70 \\ 
\hmol	      & 2.122/320  & 15 & 100 & 2000 & 1.03 \\
	          &            & 18 &  10 &  200 & 0.65 \\
	          &            & 20 &  30 & 1800 & 0.78 \\
{\it H}	    & 1.625/3100 & 15 &   5 &  500 & 1.13 \\
{\it K}$_c$ & 2.270/350  & 15 & 100 & 2000 & 1.05 \\ 
            &            & 18 &  20 &  200 & 0.75 \\   
{\it H}$_c$ & 1.570/260  & 18 &  50 &  500 & 0.85 \\
\hline
\end{tabular}
\end{table*}

\begin{figure*}
\rotatebox{-90}{\includegraphics[width=0.47\hsize,clip]{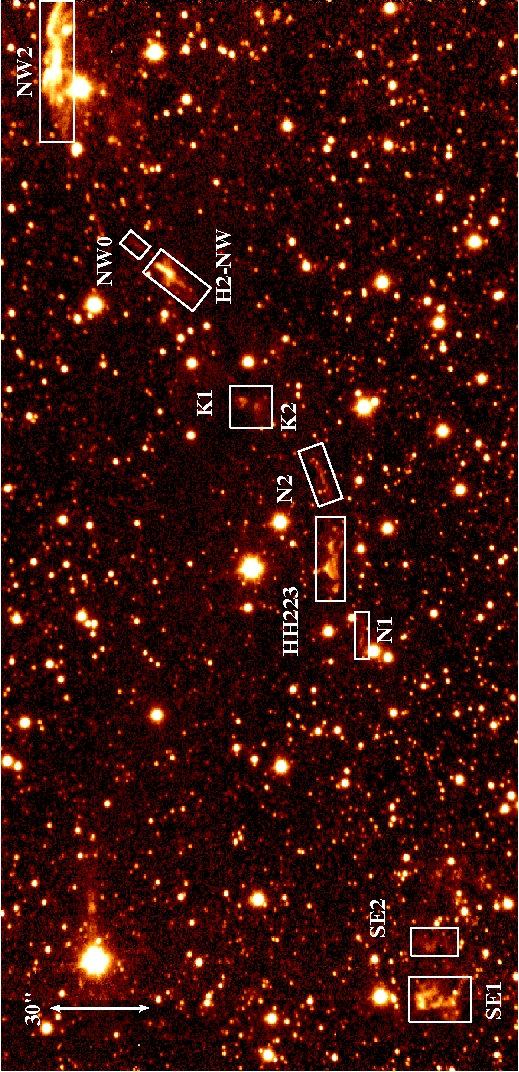}}
\caption{The L723 field imaged with LIRIS through the H$_2$ 2.122 $\mu$m line
filter (continuum is not subtracted). The emission structures of  the
HH~223 outflow have been labeled.  North is up and East is to the left. 
\label{ph2}} 
\end{figure*}

\section{Observations, data reduction and calibration}

Narrow-band images were obtained through filters centred on the \fe\  and
\hmol\ lines which is adequate to detect the emission from shocked ionized
and
molecular gas respectively.  Additional images were obtained through the
broad-band {\it H} filter and the narrow-band emission-line free filters
{\it H}$_c$ and  {\it K}$_c$, which are useful to subtract the continuum from
emission-line filters.  Observations were made on  2006 July with the
instrument LIRIS  \citep[Long-Slit Intermediate Resolution Infrared 
Spectrograph;][]{aco03} at the  4.2~m Williams Herschel Telescope (WHT) of
the Observatorio del Roque de los Muchachos (ORM, La Palma, Spain). Details
about the observations are listed in Table \ref{tlog}. LIRIS is equipped with a
Rockwell Hawaii $1024\times1024$ HgCdTe array detector. The spatial scale
is $0\farcs25$~pixel$^{-1}$, giving an image field of view (FOV) of $4\farcm27 \times
4\farcm27$.  The observing strategy was a 5-point dithering pattern. Given the
elongated  morphology of the target we used a E-W offset three times larger than
that used along the N-S direction. 

The data were processed with the  package {\it lirisdr} developed by the LIRIS
team within the  {\small IRAF} environment\footnote{{\small IRAF} is distributed
by the National Optical Astronomy Observatories, which are operated by the
Association of Universities for Research in Astronomy, Inc., under cooperative
agreement with the National Science Foundation.}. The reduction process includes
sky subtraction,  flat-fielding, correction of geometrical distortion, and
finally a combination of frames using the common ``shift-and-add''
technique.  This final step consists of dedithering and co-adding frames
taken at different dither points to obtain a mosaic for each filter.  The
resulting mosaic covers a FOV of $\sim$ 5 arcmin$^2$ (see Fig.~\ref{ph2}). Note
that narrow-band images taken with LIRIS are commonly
affected by fringing. An efficient method to correct this effect is to create a
superflat image by a median  combination of all frames corresponding to 
a filter and
rejecting pixels that contain bright stars.     

Continuum-subtracted images of the emission in the \fe\ and H$_2$ lines  were
obtained by removing the off-line emission using the images acquired  through
the {\it H}$_c$ and {\it K}$_c$ filter respectively. For each pair of {\it H}
and {\it K} images, a flux scaling factor was derived by comparing the counts of
several stars in both images and then subtracting the registered and scaled
images. In some cases the images with best seeing were degraded by
convolution with a Gaussian in order to match the width of the point spread
function (PSF).

The astrometric calibration of each  final image was made with the aim of
properly comparing the structures at near-infrared and optical
wavelengths, and with the position of the radio continuum sources found in the
field.  To do that, the near-infrared images were registered using the
coordinates from the 2MASS All Sky Catalogue of ten field stars that are well
distributed over the observed field. The rms of the transformation was
$0\farcs04$  in both coordinates.

\begin{figure}
\rotatebox{0}{\includegraphics[angle=-90,width=\hsize,clip]{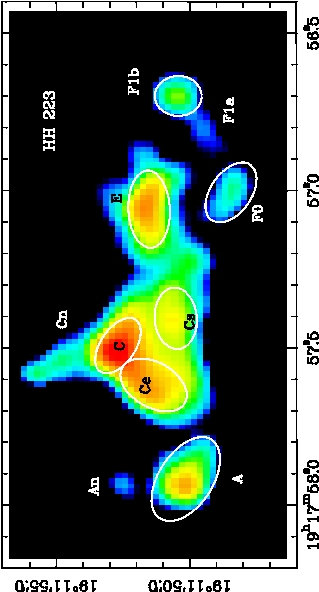}}
\caption{Close-up of the continuum-subtracted H$_2$ image showing  the structure
of  the HH 223 emission at the centre of the L723 field. The  emission
substructures referred to in Col. 1 of Table \ref{th2_c} have  been labeled. The
white ellipses enclose the regions  which correspond to the emission-line
fluxes  referred to in Cols. 3 and 4 of Table \ref{th2_c}. 
\label{closeup1}} 
\end{figure}

\begin{figure}
\rotatebox{0}{\includegraphics[width=\hsize,clip]{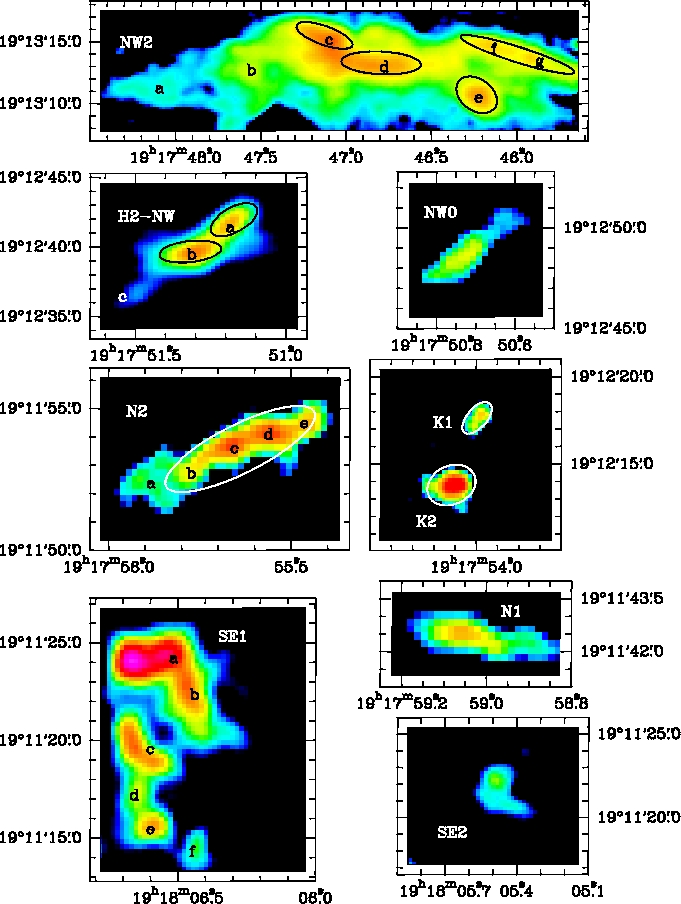}}
\caption{Close-up of the continuum-subtracted H$_2$ image showing the other
H$_2$ line-emission nebulae of the HH~223 near-infrared flow  (see Table 
\ref{th2_c} for identification). The ellipses are the same as in Fig.\ \ref{closeup1}.
\label{closeup2}} 
\end{figure}

In order to estimate the brightness of the detected emission-line structures we
determined a photometric zero point for each filter.  For this purpose we
obtained instrumental magnitudes in each of the final images by using the
image-analysis tools of the GAIA interface facility\footnote{GAIA is a
derivative of the Skycat catalogue and image display tool, developed as part of
the VLT project at ESO.}. Firstly, stars in the images were detected with the 
task {\it object detection} and then the zero-point was determined through a comparison
with the magnitudes from the 2MASS catalogue. The accuracy reached was better
than 0.1 mag in all the filters. The limiting magnitudes (at a 3$\sigma$
level) found for our images were  20.5 mag~arcsec$^{-2}$ in {\it H}$_c$  and
18.5 mag~arcsec$^{-2}$ in {\it K}$_c$.  Conversion from magnitude to flux
density was made by using the flux of Vega, which was properly interpolated taking into
account the narrow-band filter response. Then, the flux
was measured for each substructure
that could be isolated within the nebular  emission features. We
defined an ellipse encircling  the emission of the substructure  to
include signal down to a $\sim$ 3$\sigma$ limit by means of the {\it Aperture
photometry} task in GAIA (see Figs.~\ref{closeup1}, \ref{closeup2}, and
\ref{closeupfe}). The background emission was extracted from an ellipse with
the same area but encircling a region close to the substructure, which was 
supposedly
free  from line emission.  For close substructures, which are difficult to
isolate,  we  preferred to enclose  them all together within a single ellipse
and evaluate the total flux inside. Fluxes are reported in Tables \ref{th2_c}
and \ref{tfe}.

\section{Results}
\subsection{Outflow morphology in H$_2~\upsilon$ = 1--0}

\begin{table*}[htb]
\scriptsize
\caption[ ]{Positions  and photometry of the H$_2$ features detected in the
L723 field}
\label{th2_c}
\begin{tabular}{lccccc}  
\hline\hline
 &\multicolumn{2}{c}{Peak Position}&\multicolumn{2}{c}{Photometry}&\\
 &$\alpha_{2000}$ & $\delta_{2000}$  & Integrated area  & H$_2$ line flux & \\
H$_2$ feature &{(h m s)}&{($^\circ$ $\arcmin$ $\arcsec$)}&(arcsec$^2$)&($10^{-15}$
 erg~cm$^{-2}$~s$^{-1})$&Notes\\
\hline 
SE1 (V83) region&             &             && & (1),(4)\\
\phantom{0}a   & 19 18 06.63  &+19 11 24.6  && & \\
\phantom{0}b   & 19 18 06.47  &+19 11 23.0  && & \\
\phantom{0}c   & 19 18 06.65  &+19 11 19.3  && & \\
\phantom{0}d   & 19 18 06.67  &+19 11 17.1  && & \\
\phantom{0}e   & 19 18 06.61  &+19 11 15.4  && & \\
\phantom{0}f   & 19 18 06.44  &+19 11 14.0  && & \\
\hline
 SE2           & 19 18 05.48  &+19 11 22.1  && & (1),(4)\\
\hline
N1             & 19 17 59.04  &+19 11 42.3  &3.77  &0.51$\pm$0.08 & (1),(5)\\
\hline
HH~223         &              && & & (2),(6)\\
\phantom{0}An  & 19 17 57.96  &+19 11 52.5  && & (3)\\
\phantom{0}A   & 19 17 57.96  &+19 11 50.1  &7.30   &4.16$\pm$0.13 & (2)\\
\phantom{0}Cn  & 19 17 57.55  &+19 11 54.8  && & (3)\\
\phantom{0}Ce  & 19 17 57.64  &+19 11 51.1  &4.68   &6.85$\pm$0.10 & (3)\\
\phantom{0}C   & 19 17 57.51  &+19 11 52.8  &3.14   &5.58$\pm$0.07 & (2)\\
\phantom{0}Cs  & 19 17 57.42  &+19 11 50.6  &3.74   &2.78$\pm$0.10 & (3)\\
\phantom{0}E   & 19 17 57.04  &+19 11 51.5  &4.64   &4.04$\pm$0.12 & (2)\\
\phantom{0}F0  & 19 17 56.98  &+19 11 48.0  &3.94   &0.85$\pm$0.07 & (2)\\
\phantom{0}F1a & 19 17 56.78  &+19 11 49.3  && & (2)\\
\phantom{0}F1b & 19 17 56.67  &+19 11 50.3  &2.69   &1.26$\pm$0.08 & (2)\\
\hline
H2-N2          &              &             &10.32  &0.59$\pm$0.15 & (1),(7)\\
\phantom{0}a   & 19 17 55.87  &+19 11 52.3  && &    \\
\phantom{0}b   & 19 17 55.78  &+19 11 52.6  && &    \\
\phantom{0}c   & 19 17 55.66  &+19 11 53.8  && &    \\
\phantom{0}d   & 19 17 55.55  &+19 11 54.2  && &    \\
\phantom{0}e   & 19 17 55.42  &+19 11 54.6  && & (3)\\
\hline
K2             & 19 17 54.10  &+19 12 13.6  &7.15   &1.54$\pm$0.11 & (3)\\
\hline
K1             & 19 17 53.96  &+19 12 18.1  &2.81   &0.53$\pm$0.07 & \\
\hline
H2-NW          &              &             && & (3)\\
\phantom{0}a   & 19 17 51.17  &+19 12 42.0  &4.24   &12.6$\pm$0.10 &\\
\phantom{0}b   & 19 17 51.30  &+19 12 39.6  &4.23   &14.2$\pm$0.08 &\\
\phantom{0}c   & 19 17 51.51  &+19 12 36.3  && &\\
\hline
H2-NW0         & 19 17 50.68  &+19 12 49.1  && & (8)\\
\hline
H2-NW2         &              &             && & (9)\\
\phantom{0}a   & 19 17 48.11  &+19 13 11.4  && &\\
\phantom{0}b   & 19 17 47.56  &+19 13 14.4  && &\\
\phantom{0}c   & 19 17 47.09  &+19 13 14.8  &7.11   &43.4$\pm$0.16 &\\
\phantom{0}d   & 19 17 46.80  &+19 13 13.1  &11.51  &37.3$\pm$0.19 &\\
\phantom{0}e   & 19 17 46.20  &+19 13 10.3  &8.97   &33.8$\pm$0.21 &\\
\phantom{0}f   & 19 17 46.11  &+19 13 15.0  &11.75  &17.3$\pm$0.22 &(10)\\
\phantom{0}g   & 19 17 45.88  &+19 13 13.6  && &\\
\hline
\end{tabular}
\begin{list}{}{}{}{}{}
\item[] Notes:\\
(1) With  counterpart in H$\alpha$;
(2) with  counterpart in H$\alpha$ and \sii;
(3) without optical counterpart;
(4) position of the apex;
(5) position of the peak intensity;
(6) this emission has been labeled as its optical counterpart; 
    the region between knots C and
    F corresponds to the H$_2$ emission labeled as H2-SE
    by \cite{pal99};
(7) knots b to e have been included in the flux value;    
(8) first detection in this work;
(9) weak H$\alpha$ counterpart; detected in {\it K'} by  \cite{hod94};
(10) knots f and g have been  included in the flux value.
\end{list}
\end{table*}

Figure\ \ref{ph2} displays an image of the L723 field through the \hmol\ 
narrow-band filter that shows the large-scale  emission of the HH~223  outflow. 
As can be seen from the figure, the emission spreads over a length of 
$\sim5\farcm5$ ($\sim0.5$ pc for a distance of  300~pc) from the southeast  to
the northwest. It consists of  several groups of nebulae  of different sizes and
complexity, distributed following a S-shaped  pattern that lies projected onto
the lobes of the east-west CO outflow.  The morphology outlined at large scale
is then suggestive of a parsec-scale H$_2$ outflow.  

The large-scale morphology of the H$_2$ emissions agress well with that
of the  atomic (H$\alpha$, \sii) emissions in the loci where  both optical and
near-infrared emissions are detected. We found the  H$_2$ counterparts of
the H$\alpha$ nebulae HH223-SE1,  -SE2, -N1, -N2, and of the largest 
emission, HH~223 \citep{lop06}. Additional H$_2$ emission nebulae without
visible counterpart appear towards the northwest of HH~223. Some of them   were
previously detected by \citet{pal99} (\eg\ labeled in their work as K1, K2 and
H$_2$-NW). Our new images give a wider coverage of the HH~223 near-infrared
outflow. They include new faint emissions (\eg\  HH223-NW0, towards the
northwest of the field) and the   filamentary  emission {\it K'} reported by 
\citet{hod94}, which has a   H$\alpha$  counterpart (HH223-NW2) weaker and less
extended.   In addition, we were able to distinguish the emission contributions
from the continuum and from  the H$_2$ line. The continuum-subtracted H$_2$
image shows several emission spots that appear to be extended nebulae
surrounding embedded stars, whose physical association with the stars cannot be
established  from the present data. In contrast, most of the filamentary
emission features come from ``true'' line emission that appear complex and knotty, with
several condensations embedded inside  a more diffuse nebular emission. 

The ``true'' H$_2$ line emission nebulae along the HH~223 near-infrared 
outflow  were identified from the continuum-subtracted H$_2$ image,  and those
with a spatially-extended  structure have been resolved in smaller
substructures.   In this work, the H$_2$ nebulae have been labeled according to  the
nomenclature of  \citet{pal99}, but we expanded it with  the newly found
 near-infrared features.
There is an exception concerning to the  emission of
$\sim30\arcsec$ length towards the centre of the field (\ie\  HH~223),
where we keep the nomenclature   established in  \citet{lop06}  for the
H$\alpha$ and \sii\ knots. Similar to what we did for 
 the  H$\alpha$ and \sii\
line-emission features, we added the identification label  {\it HH~223-} 
to all the H$_2$  nebulae found projected onto the east-west CO outflow, since
they most probably form part of  the same near-infrared/optical outflow. 

The H$_2$ emission structures identified from our image are listed in Table
\ref{th2_c}  (Col. 1), and their positions are given in Cols. 2 and 3. The
H$_2$ line flux of the nebulae,  integrated over the area of Col. 4, is
given in Col. 5. Figure \ref{closeup1} displays a close-up of  HH~223, showing
the morphology of the H$_2$ emission with detail. Figure \ref{closeup2} shows
close-up of several extended nebulae (those  containing smaller substructures)
along the HH~223 near-infrared outflow.

\begin{table*}
\caption[ ]{Positions and photometry of the \fe\  features detected in 
the L723 field}
\label{tfe}
\begin{tabular}{lcccc}
\hline\hline
 &\multicolumn{2}{c}{Peak Position}&\multicolumn{2}{c}{Photometry}\\
 &$\alpha_{2000}$ & $\delta_{2000}$  & Integrated area  & \fe\ line flux  \\
\fe\ feature  &{(h m s)}&{($^\circ$ $\arcmin$ $\arcsec$})&(arcsec$^2$)
&($10^{-15}$ erg~cm$^{-2}$~s$^{-1})$\\
\hline 
 HH~223$^{(1),(2)}$&          &           &&   \\
\phantom{0}A   & 19 17 57.99  &+19 11 50.4&4.68  &$5.95\pm0.10$\\
\phantom{0}B   & 19 17 57.73  &+19 11 49.7&4.58  &$3.39\pm0.10$\\
\phantom{0}C   & 19 17 57.45  &+19 11 52.9&5.83  &$3.64\pm0.09$\\
\phantom{0}D   & 19 17 57.26  &+19 11 52.2&3.68  &$1.16\pm0.09$\\
\phantom{0}E   & 19 17 57.05  &+19 11 51.8&3.68  &$0.71\pm0.07$\\
\phantom{0}F0  & 19 17 56.96  &+19 11 47.2&2.28  &$0.14\pm0.25$\\
\phantom{0}F1a & 19 17 56.77  &+19 11 49.6&1.94  &$0.28\pm0.06$\\
\phantom{0}F1b & 19 17 56.68  &+19 11 50.3&2.18  &$0.37\pm0.07$\\
\hline
\end{tabular}
\begin{list}{}{}{}{}
\item[] Notes:\\
(1) The knots were labeled as its optical counterparts. Knot B has not
been detected in H$_2$;
(2) 
note the displacements between the \fe\ and H$_2$ emission peak positions
(see text). 
\end{list}
\end{table*}

We will briefly describe the main emission nebulae below, beginning with HH~223.
The rest of  H$_2$ emission nebulae are listed from southeast to
northwest.

\begin{description}

\item[\bf HH~223:]
The knotty, undulating structure of $\sim22\arcsec$ in length 
found towards the centre of the field, $\sim1\arcmin$ southeast of the radio
continuum system VLA~2. The H$_2$ feature spatially coincides  with the
H$\alpha$ ``linear emission feature'' reported by \citet{vrb86}, which
corresponds to HH~223 in the Reipurth Catalogue of Herbig-Haro objects
\citep{rei94}. Part of this structure (from HH~223-C to HH~223-F) corresponds to
the H2-SE emission of \citet{pal99}. The field  imaged by these authors
covers up to the eastern edge of HH~223-C, so that  HH~223-A was not
mapped. Since the  overall morphology and  size of  HH~223  is similar for the
optical and near-infrared emissions and to facilitate  the
comparison between them, we adopt the  nomenclature for the knots  of
\citet{lop06}, instead of using the  H2-SE label given in \citet{pal99}
for the part of the H$_2$ emission mapped by them.  In spite of the
similarity between the optical and near-infrared emissions, a more detailed
comparison shows differences that will be discussed later.  

\item[\bf HH~223-SE1 -SE2:]
A group of H$_2$ nebulae $\sim2\arcmin$ southeast of  HH~223-A,  the
counterparts of the H$\alpha$  nebulae HH~223-SE1  (also associated with the V83
nebula of \citealp{vrb86}) and HH~223-SE2, reported by \citet{lop06}.  
HH~223-SE1, unresolved in H$\alpha$, consists of several  arc-shaped  structures
of different sizes (SE1-a to -f) in the H$_2$ image.   HH~223-SE2 appears
in H$_2$ as a faint, bow-shaped structure (labeled SE2 in Table \ref{th2_c} and
Fig.\ \ref{closeup2}).

\item[\bf HH~223-N1:]
Faint, filamentary S-shaped emission that extends over $\sim11\arcsec$ in
length, $\sim15\arcsec$ southeast of HH~223-A. Only the brightest eastern 
part of it, of $\sim5\arcsec$ in length was barely detected in H$\alpha$.

\item[\bf HH~223-H2-N2:]
Faint filamentary H$_2$ feature  $\sim 30 \arcsec$  northwest of HH~223-A and
extending over $\approx 15\arcsec$ towards the northwest, projected onto  the
blueshifted  CO outflow lobe. This emission was barely detected in the image of 
\citet{pal99}. Our H$_2$ image shows several brightness emission enhancements
(labeled -b to -e in Table \ref{th2_c} and Fig.\ \ref{closeup2}).  No visible
counterpart was detected for this emission, except for its faint  eastern part
(H2-N2-a), which coincides with the H$\alpha$ emission HH~223-N2.

\item[\bf HH~223-K2, -K1:]
These two emission nebulae, $\sim60\farcs5$ (K2) and 
$\sim64\arcsec$ (K1) northwest of HH~223-A, are the nebulae closest
to the radio continuum multiple system VLA~2. They were first reported by 
\citet{pal99} as two low-brightness H$_2$ patches.  
K2 appears in our images as a quite  compact knot of $\sim3\arcsec$  in
diameter. Its emission mostly arises from  the H$_2$ line, as shown in the
H$_2$ continuum-subtracted frame (see panels in Figs.\ \ref{closeup1} and
\ref{k2k1}). 
The emission of K1  has contributions from both continuum and line.
In the the H$_2$   continuum-subtracted frame, K1  shows an elongated shape
extending $\sim3\arcsec$ along a direction with a position angle
$\sim130^\circ$ (see panel of Fig.\ \ref{closeup1}). In contrast its
morphology is somewhat different in the H$_2$ frame keeping the continuum:
the K1 emission extends in an emission tail towards its eastern side and,
in addition, K1 appears brighter than K2. 

\item[\bf HH~223-H2-NW, -NW0:]
HH~223-H2-NW is the emission of $\sim12\arcsec$ length,  found
$\sim1\rlap.'5$ northwest of HH~223-A. Its shape 
follows a smooth undulating pattern,  
with at least  three substructures enclosed in a more diffuse emission. North of
HH~223-H2-NW, an additional filamentary emission of $\sim2\arcsec$
length has been detected in our images. It  has been labeled HH~223-NW0 to
distinguish it from the rest of the HH~223-NW emissions found in the
H$\alpha$ and
H$_2$ images, 
which are located few arcmin from HH~223. 

\item[\bf HH~223-H2-NW2:]
The bright, filamentary H$_2$ emission, with several
substructures inside, which is seen towards the northwestern edge of the field.
The emission coincides with  a {\it K$^\prime$} linear nebula reported by 
\citet{hod94}. Part of the HH~223-H2-NW2 structure is the counterpart of the
faint H$\alpha$ nebula HH~223-NW2.
\end{description}

\subsection{\fe\ imaging}

\begin{figure}
\rotatebox{0}{\includegraphics[angle=-90,width=\hsize,clip]{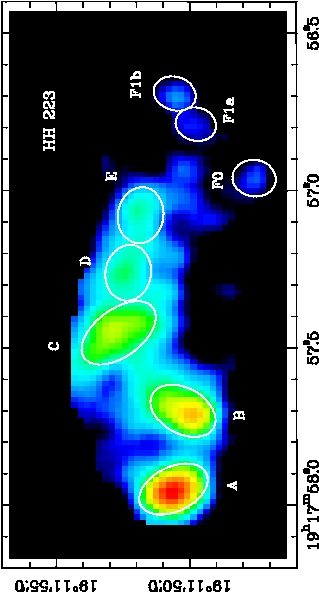}}
\caption{Close-up of the continuum-subtracted \fe\ image showing  HH~223. 
The knotty substructures referred in Table \ref{tfe} have 
been labeled on the figure. The ellipses are the same as 
in Figs.\ \ref{closeup1} and \ref{closeup2}.
\label{closeupfe}} 
\end{figure}

Additional images of the L723 field were obtained in the {\it H}-band (see
Table~\ref{tlog}). After subtracting the {\it H}$_c$ from the \fe\ + continuum
images, significant  emission from the \fe\ line was only detected at
the location of  HH~223. The  structure  of the \fe\
emission consists of  several knotty condensations inside a lower-brightness
nebular emission (see Fig. \ref{closeupfe}). In Table \ref{tfe} we list the
\fe\ condensations of HH~223 (Col. 1), their peak position
coordinates (Cols. 2, 3) and the \fe\ line flux (Col. 5), integrated within
the area given in Col. 4.  As can be seen from the figure, the \fe\ emission 
of HH~223  extends from east to west over a similar length than the
emissions in the H$_2$ and H$\alpha$ lines. The overall morphologies are also
similar. However, some significant differences are found when these emissions
are compared in detail. We will discuss this point in the next section. 

Emission  from the \fe\ line  was not detected out of HH~223.  Emission from
this line has been frequently found close to  the location of embedded
young stellar objects (YSOs), and traces  the  innermost part of the jet that is
accelerated near the driving  source (see, \eg\ \citealp{Rei00}). Hence, one
would expected to find \fe\ emission associated with  K1, the  H$_2$ nebula
found closest to the radio continuum system  VLA~2. However, no emission
was detected in our images around K1. Likely, the extinction is  high enough to
prevent the detection, provided that the
degree of ionization of the outflow is low enough to produce only faint \fe\ 
emission.

It is not reasonable to expect emission  from the \fe\ line at  regions having
H$\alpha$  emission and no counterpart in the \sii\ lines (like HH~223-SE or
HH~223-NW2). This is because of  the 1.644 $\mu$m \fe\ and \sii\ lines arise
from similar excited atomic levels, and both species have similar ionization
potentials.

\section{Discussion}
\subsection{Comparing  the optical and near-infrared emissions of the 
HH~223 outflow}

\begin{figure*} 
\rotatebox{-90}{\includegraphics[height=\hsize,clip]{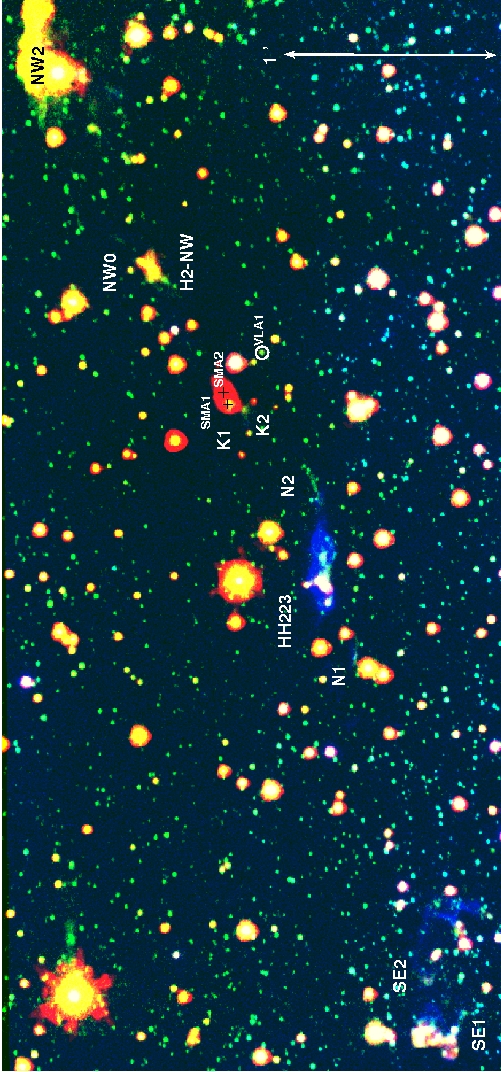}}
\caption{Composite image of the L723 field using a H$\alpha$ (blue), H$_2$  2.122 $\mu$m (green) and IRAC 4.5 $\mu$m (red) colour coding
to show the HH~223 outflow emissions. The positions of the radio continuum sources 
SMA1 and SMA2 have been marked by plus signs. The near-infrared counterpart of
the radio source VLA~1 has been marked by a circle. The continua were not subtracted 
to facilitate the alignment of all  images using the field stars.
\label{imagecomp1}} 
\end{figure*}

It is interesting to compare the morphology of the HH~223 outflow emission for 
different  spectral ranges. This can be done from the available images of the 
the L723 field that include the outflow.  

At optical wavelengths, we   use the deep narrow-band H$\alpha$ and \sii\
images acquired with the 2.6-m Nordic Optical Telescope (NOT) \citep[see ][ for
details of the observations]{lop06}.  In addition to the near-infrared images in
the {\it H} and {\it K} bands discussed in this work,  we  use an image 
of the Spitzer Infrared Array Camera (IRAC) through its channel 2 (centred at
4.5 $\mu$m).  For the typical conditions of low-mass outflows, emission from
several rotational transitions lines of H$_2$  should be expected  within  the
$\approx$ 4--8 $\mu$m wavelength range. It is well-known that the IRAC channel 2 is
the best suited in detecting shock-excited H$_2$ features, in contrast with the IRAC  channels 3 (5.8 $\mu$m) and 4 (8.0
$\mu$m). This is because  channel 2 is the band that is
less contaminated by extended emission arising from polycyclic aromatic
hydrocarbon (PAH). The IRAC channel 2 image of the  L723 field was obtained 
from the Spitzer Science
Archive using the {\small Leopard} software. 
The image used here  corresponds to the
post-basic calibrated data (pbcd) obtained in the project P00139 {\it From
Molecular Cores to Planet}  (PI N. Evans).

The original images of the  bands  we  compared have a different spatial scale
and FOV. In order to  do more meaningful comparisons, all  images were
transformed to reach the same magnification. First, we extracted a subimage of
the IRAC frame after rotating the original pbcd frame by the appropriate angle
to be oriented in the standard way. Then, we identified six stars that are 
common to all
images and are  well distributed in the field. These stars were used to
register the images  for converting all of  them into a common reference system
with the {\sc iraf} tasks {\sc geomap} and {\sc  geotran}.  All the final
five frames (\ie\ H$\alpha$, \sii, \fe, H$_2$ at 2.122 $\mu$m  and at 4.5
$\mu$m) used here to compare the HH~223 outflow emissions have a
spatial scale of 0\farcs6~pixel$^{-1}$.

\subsubsection{Atomic and molecular hydrogen emissions}

Figure \ref{imagecomp1} displays a composite image of the L723 field, where the 
emission from  atomic (H$\alpha$) and molecular (H$_2$ ~2.122~$\mu$m and
4.5~$\mu$m) hydrogen lines  appear superposed.  The figure reveals that the
large-scale morphology of the HH~223 outflow is very similar.   However, slight
differences in several structures can be appreciated  when  the hydrogen
emission  from each  band is compared in detail.  

The HH~223-SE1, SE2 and N1 nebulae at  the southeastern side of the HH~223
outflow are detected  in the three bands, and they are spatially coincident. However,
HH~223-SE2 shows a   bow-shaped  morphology in the two H$_2$ emission bands,
while this shape is not as clearly outlined in  H$\alpha$, where the emission
appears more extended and filamentary.

To the northwestern side of the outflow, the HH~223-H2-NW and  H2-NW2 nebulae
appear bright and with a very  similar morphology in the two H$_2$ emissions,
while in  H$\alpha$ H2-NW was undetected, and  H2-NW2  was barely   
detected and appears shorter. These differences could be due to the
extinction, since both structures are located within a region that appears
highly obscured at optical wavelengths. 

The faint filamentary feature HH~223-N2 detected at 2.122 $\mu$m has neither
H$\alpha$ nor 4.5 $\mu$m counterparts. The lack of H$\alpha$ emission a this
location should be  expected because of the high visual extinction. In
contrast,  we believe  that  the lack of counterpart at 4.5 $\mu$m might be
better attributed  to the weakness of the emission,  which is below the detection
limit of the IRAC frame. 

HH~223, at the centre of the field, shows a  morphology and a spatial brightness
distribution  closely coincident for the emissions in  molecular hydrogen at 
2.122 $\mu$m and 4.5 $\mu$m. Some differences appears however when the H$_2$
and  H$\alpha$ emissions are compared in detail. One difference is that the
low-brightness emission surrounding the knots appears more extended in H$\alpha$
than in H$_2$. The knots appear better delimited, with their edges  sharper
in H$\alpha$ than in H$_2$. Another more relevant difference is the lack of
counterparts in H$_2$ for knot B, which is very bright in  H$\alpha$, and for
knot D, which cannot be resolved as a knotty structure within the low-brightness
emission, in contrast to what happens in  H$\alpha$. Finally, several faint
knotty structures were identified in the H$_2$ 2.122 $\mu$m line without a
counterpart in H$\alpha$ (\ie\ those labeled An, Cs, Ce and F0, this last
belonging to the F filament). An additional low-brightness filamentary emission
in H$_2$ 2.122 $\mu$m, without counterpart in H$\alpha$, was identified
extending $\sim$~3\arcsec\ from knot C to the northeast.  The variations of the
physical conditions and, in particular,  of the gas kinematics  through   HH~223
seem to be  mainly responsible for the differences found between the
atomic and the molecular hydrogen emissions. We will return  to this point
later.

No H$\alpha$ counterparts were detected at the positions of the
HH~223-K1 and K2 nebulae. This should be  expected because of   the high visual
extinction in this region, where the radio continuum sources are located.
In contrast, counterparts  at 4.5 $\mu$m for both HH~223-K1 and K2 are
detected in the IRAC frame.  Note however that at 2.122 $\mu$m the emission
of HH~223-K2 appears  rather  more extended towards the southeast, away from the
radio continuum sources, than at 4.5 $\mu$m.

Concerning HH~223-K1, the emission appears significantly more extended at 4.5
$\mu$m than at 2.122 $\mu$m.  The emission at 4.5 $\mu$m  spreads towards the
northwest, beyond its  counterpart at   2.122 $\mu$m, and encompasses the region
where  the two radio continuum sources (SMA1 and SMA2) are located. In
contrast,  as can be seen in Fig. \ref{imagecomp1}, the emission of K1 at
2.122 $\mu$m does  not reach the position of SMA2
(see also Fig.\
\ref{k2k1}). The extinction is probably high enough to prevent its
detection in the {\it K} band  around this position. Furthermore, as in the case of the emission associated with K1 in
the {\it K} band,  its counterpart at 4.5 $\mu$m could have continuum
contribution from  heated material of the wall cavity opened by the outflow,
which can be traced closer  to the radio continuum sources at this 
wavelength.

\subsubsection{ \sii\ and \fe\ emissions} 

\begin{figure}
\rotatebox{-90}{\includegraphics[height=\hsize,clip]{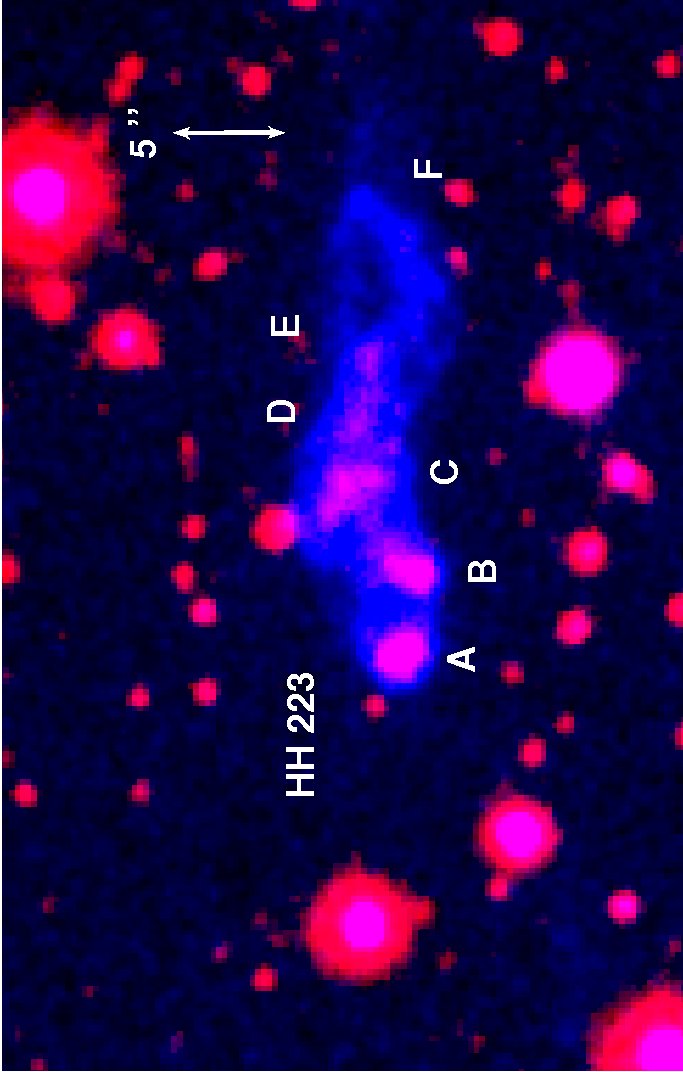}}
\caption{Composite image of HH~223  using a \sii\ 6716,31 \AA\ (blue) and \fe\
1.644~$\mu$m  (red) colour coding (continua were not subtracted).  North is up
and East is to the left.
\label{imagecomp2}} 
\end{figure}

The near-infrared \fe\ 1.644 $\mu$m emission, when detected, appear   spatially
coincident with the \sii\ $\lambda$~6716, 31 \AA\ emission.  Both emissions
usually  have a very similar morphology (see \eg\ \citealp {Rei00,Dav03}). Also,
the differences between their detailed structures are much  smaller than the
differences found between the \fe\ and H$_2$ emissions.  This is because the
\sii\  and \fe\  lines arise from atomic levels of similar excitation energy and
ionization potential. Thus,  it is expected  the emission from both lines to be
spatially coincident, tracing the shock-ionized  gas component of the outflow.  
However, the relative brightness distribution for both emissions could vary. This
is  not only owing  to differential extinction effects, but also because of the
higher \fe\  critical density ($\simeq$ one order of magnitude above the \sii\ 
critical density). Hence, the \fe\ emission requires a higher pre-shock gas
density and/or a faster shock velocity to reach an intensity comparable to the
\sii\ emission intensity.

Figure \ref{imagecomp2} shows a close-up of the L723 field including 
HH~223 from knot A to filament F, where the emissions from  \sii\
6716,31 \AA\ (in blue)  and \fe\ 1.644 $\mu$m (in red) have been  superposed. 
The knots detected in the \fe\ line (see also Fig. \ref{closeupfe})  have the
corresponding \sii\ knotty counterpart as would be expected,  although the knots
appear more compact in \fe\ than in \sii.  Moreover,  the emission  coming from
the lower-brightness nebula surrounding the knots  is much fainter in \fe\ than
in \sii.

Some clues on the spatial distribution and on the relative strength of both
emissions are obtained from the kinematics and physical conditions  through 
HH~223,  derived in \citet{Lop09} from long-slit optical spectroscopy. 

For example, the line-ratio diagnosis shows that knot A (one of the  brightest in
\fe) is the densest,  most excited and highly ionized knot (indeed, the
electron density derived at several positions inside knot A fall above the \sii\
critical density for  collisional de-excitation).  In contrast, the electron
densities derived in the  low-brightness nebula as well as in the fainter
HH~223 knots (D to F), are up to one order of  magnitude lower. In addition, 
the excitation and ionization measured in these knots (D to F) are visibly 
lower than in knot A. Hence, the excitation and density conditions in knot A
could explain its high \fe\ emission as compared with the rest of the knots.
This is consistent with the results derived by \citet{Nis02} and \citet{Nis05}
in several HH objects with a more accurate diagnostic from combined
optical/near-infrared spectroscopy. These authors were able to detect a density
stratification from ratios of forbidden lines with different critical
densities (such as \sii\ and \fe) as well as variations of the spatial
distribution  of the ionization/excitation conditions along the jet emission. In
particular, they derived that the \fe\ emission comes from the densest and/or
highly  ionized jet regions.

On the other hand, note that knot B is  bright in \fe, but lacking of
apparent emission in H$_2$. Knot C, detected in H$_2$, has a  \fe\ emission
weaker than knot B. Concerning their conditions, from the optical spectra we
derived very similar electron densities and excitation/ionization degrees for
knots B and C. In contrast, knot B is one of the faster HH~223 knots. The radial
velocity derived for knot B  ($V_\mathrm{LSR}=-125$~\kms) is
appreciably higher (by a factor of two) than for knot C 
($V_\mathrm{LSR}=-55$~\kms).   
Thus, we interpret that the discrepancies found  between  the \fe\ and \sii\
emissions are more likely produced by the kinematics and physical conditions in the
knots than by differential extinction effects: knots denser and/or faster are
stronger emitters in the \fe\ line  than slower and/or more rarefied knots.
Furthermore, the  emission appears spatially more extended in \sii, which
gives rise to an extended low-brightness  nebula around the knots, than in \fe.
 This could be explained because we are detecting \fe\ emission only from 
the higher
density material of the knots. Thus,  the
shape of the overall \fe\
emission appears more compact than the \sii\ emission.
 The lack of  detection of
the H$_2$ line coinciding with the position of the fastest  knot B also 
supports this interpretation, since the dissociation of the H$_2$
molecule  can be more efficient at the  knots with higher velocity. 

\subsubsection{ \fe\ and \hmol\ emissions}

\begin{figure}
\rotatebox{-90}{\includegraphics[height=\hsize,clip]{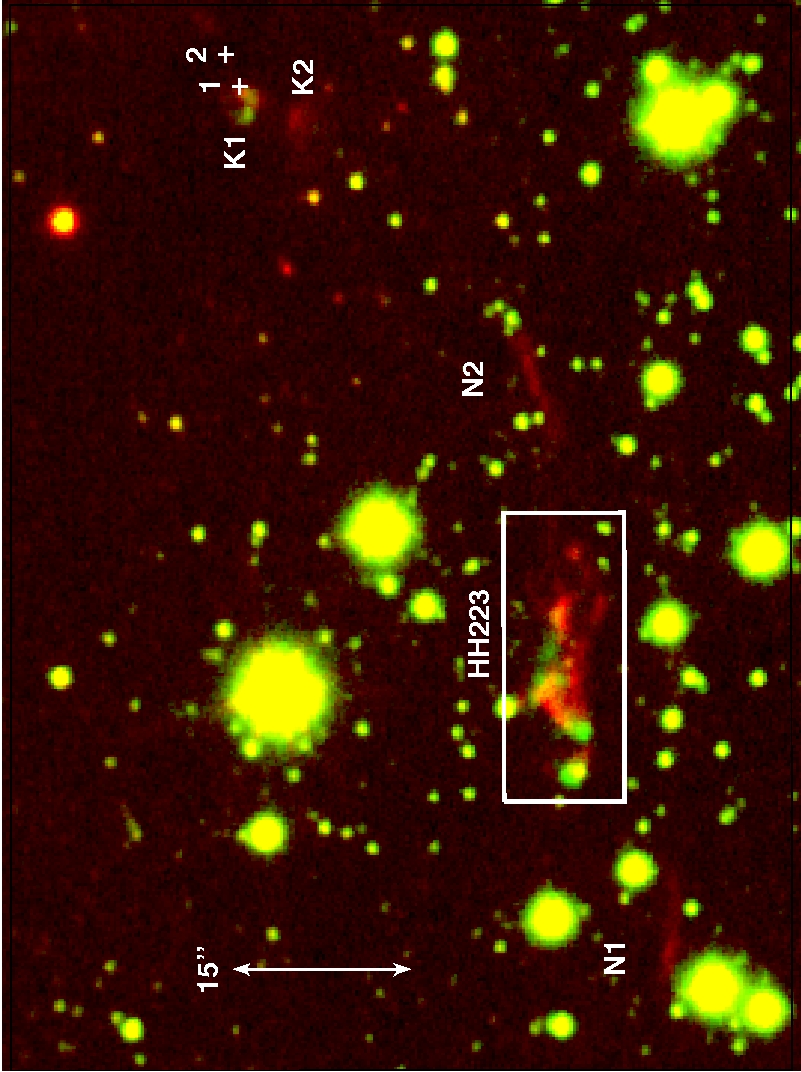}}
\rotatebox{0}{\includegraphics[height=0.395\hsize,clip]{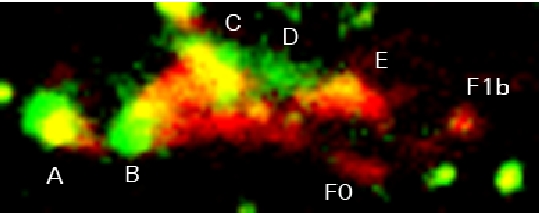}
}
\caption{\emph{Top panel}: Composite image,
 using a  H$_2$~2.122~$\mu$m (red) and \fe~1.644~$\mu$m (green) colour coding,
 of a FOV $\sim 1\farcm5 \times 1\farcm1$ of L723 showing part of the HH~223
 outflow (continua were not subtracted). Crosses mark the position of the radio continuum
sources SMA1 and SMA2, close to the H$_2$ structure K1 (see also Fig. \ref{k2k1}).
 \emph{Bottom panel}: Close-up of the $\sim 24\farcs3
\times 10\farcs2$ rectangle enclosing HH~223. Note the spatial displacements between the
\fe\ and  H$_2$ emissions in the knots. North is up and East is to the
 left.
\label{h2fe}} 
\end{figure}

Usually, the H$_2$ and \fe\ emissions from jets show different spatial
brightness distributions: the  H$_2$ emission is more diffuse and extended
than the \fe\ emission (see \eg\ \citealp{Rei00}). Here we will compare
in greater detail the emission   from both species in the HH~223
outflow. Figure \ref{h2fe} displays a composite image, where the emissions from
\fe\  1.644 $\mu$m (in green) and H$_2$ 2.122 $\mu$m (in red) appear
superposed. 

Several differences between the spatial brightness distribution of these two
emissions are appreciated. First, the shape of some knots is different.
This is  clearly seen \eg\ for knot E. The emission of this knot appears more
extended in H$_2$ and shows a bow-shaped morphology as compared to \fe, 
where the
emission is more compact.  Second, the relative intensity between the knots
is also different. Note, \eg\ that knots A and C show a similar brightness in
H$_2$, while knot C appears appreciably weaker  in the \fe\
line compared to knot A. Even more noticeable is the lack of  H$_2$ emission at the position
of knot B, which is one of the brightest in \fe.  Finally, we 
obtained reliable measurements of the position offsets between
the \fe\ and H$_2$ emission  intensity peaks: 
In knot A the \fe\ emission peaks $\sim0\farcs7$ northeast of the H$_2$
emission peak;  in knot C  the \fe\ peak is $\sim1\farcs1$
northwest of the H$_2$ peak (see also Fig.\ \ref{h2fe}, bottom panel). 

Again, we  interpret these differences  between both emissions as mostly caused
by the excitation conditions required for each line to emit, namely shock
velocity and gas density, rather than to differential extinction effects.
Further support of this interpretation comes from the anticorrelation 
 that  we found
 between the radial velocities of the knots in   HH~223,
derived from optical spectroscopy, and their relative H$_2$ brightness. 
The radial velocity decreases  for the sequence of
knots B, D, A, E and C, while the H$_2$ brightness increases for the
sequence of knots C, E, A; the emission is very weak in D and is below
detection limit in knot B where, in addition, the degree of ionization  inferred
from optical line ratios reaches the lowest value measured in all  knots.

As mentioned before, \fe\ line emission  was only detected associated
with  HH~223. The {\it H}-band emission detected around the location of K1, the
H$_2$ nebula closest to the radio continuum sources,  mainly arises from
continuum. 

Some of the lower-brightness diffuse H$_2$ emission features without \fe\
counterpart may trace  slower shocks in the ambient molecular gas with which the
supersonic jet is colliding. This was suggested to explain the nature of the
H$_2$ emission structures lacking of \fe\ counterpart that, as in our case, were
detected  along the molecular outflow lobe associated with the jet powered by
L1551 NE \citep{Hay09}.

\subsection{Near-Infrared environment of the radio continuum  multiple
system VLA 2}
 
\begin{figure} 
\rotatebox{0}{\includegraphics[width=\hsize,clip]{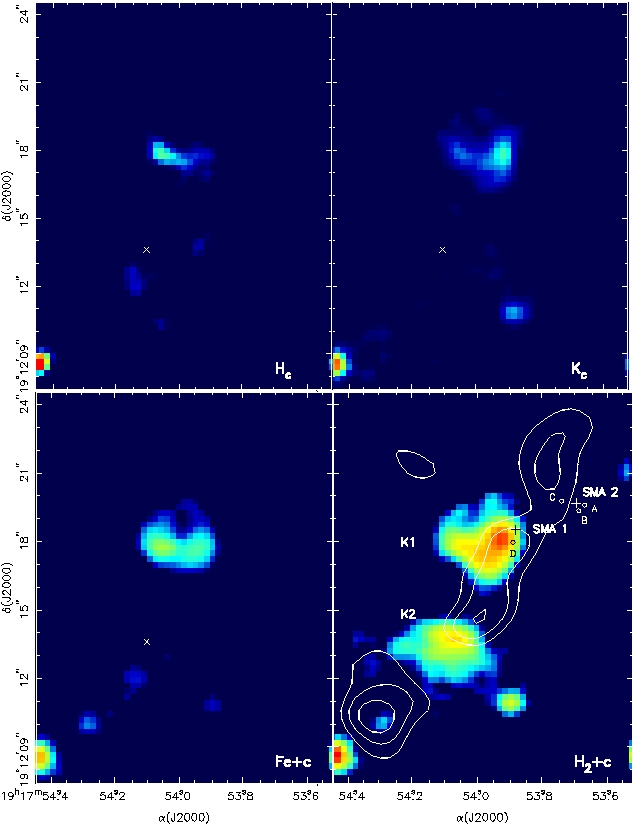}}
\caption{Close-up of the L723 field including the K2 and K1 near-infrared
structures, in the
neighbourhood of the radio continuum  multiple system VLA~2. 
In the lower-right panel, the
crosses mark the positions of the  sources SMA1 and SMA2 detected  
at 1.35~mm by 
\cite{Gir09}. The circles mark the positions of the four components 
of VLA~2 (\ie\ 
VLA~2A to D) detected at 3.6~cm and 7~mm
by \cite{Car08}. The SiO outflow reported by \cite{Gir09} has been superposed
on the image with white contours. The position of the H$_2$ emission
peak  of  K2  has been marked with a tilted cross in the other three panels.  
\label{k2k1}} 
\end{figure} 

Figure  \ref{k2k1} displays a close-up of  L723  dark cloud showing the field 
around the low-mass protostellar multiple system (VLA~2A to 2D) embedded in the
region. The  panels of the Fig.\ \ref{k2k1} display the near-infrared emission
in the {\it H} and {\it K} bands obtained from the images of this work. 

As mentioned before, \citet{pal99} reported the detection of  two faint H$_2$
nebulae, named  K1 and K2, near VLA~2. However, our images indicate that the
nature of the emission is  different for K1 and K2. 

Concerning K1, our images indicate that its emission has contributions from
both line and continuum. This can be seen in the panels of Fig.\
\ref{closeup2} and its comparison with Fig.\ \ref{k2k1}. The morphology of K1
both in the  {\it H} and {\it K} band images including continua
consists of a nearly arc-shaped nebula, with two (east and west)
brightness  enhancements of different intensity.  The emission of
the \fe\ 1.644 $\mu$m line falls below the detection limit, as revealed from the
continuum-subtracted \fe\ image. The   emission from the H$_2$ 2.122 $\mu$m line
only  contributes to the western part of K1  (the side closer to the radio
continuum sources) and  appears  elongated   in the northwest-southeast
direction (PA $\simeq 140^\circ$). Interestingly, \citet{Gir09} reported
emission of the SiO 5--4 line towards the position of the  radio continuum
sources.  The SiO emission shows an elongated morphology, reminiscent of a jet,
which also follows the northwest-southeast direction. Moreover, the H$_2$
emission peaks of K1 and K2 are in good coincidence with two SiO emission
enhancements   (note that the positions of knots K1 and K2 in Fig.~6 of
\citet{Gir09} are offset  $\sim5\arcsec$ east of the position reported in
this work). The SiO line is a tracer of shock-excited molecular gas.
Hence, it seems reasonable  to assume that both emissions,  H$_2$~2.122 $\mu$m 
and the SiO, are tracing  the same outflow. The continuum emission  from the 
eastern side of K1  may arise from heated material of a cavity wall. 

Concerning K2, which lies south of the radio continuum system, the near-infrared
emission in the {\it H}$_c$ and  {\it K}$_c$ (continuous) bands falls below the
detection limit. In contrast K2 appears quite bright, with a magnitude of
17.3, in the H$_2$ narrow-band image, and remains without appreciable
differences after subtracting the continuum, both in shape and brightness.
Hence, K2 emission comes mostly from the 2.122 $\mu$m line, which  is
tracing shock-excited gas (note that excitation by fluorescence  cannot be a
likely option because of no ionization sources, like early-type stars, are found
associated with the L723 dark cloud). Interestingly, another clump of SiO
emission is detected $\sim6\arcsec$ southeast of the K2 intensity
peak. This clump is aligned with the CO outflow direction. This 
reinforces the idea that the near-infrared (H$_2$ line) and SiO emissions
are  tracing shock-excited gas in a jet, and points towards SMA1 as the likely
source powering the SiO outflow.

\subsection{The nature of the radio continuum source VLA~1}

\citet{ang91} detected at 3.6~cm another radio continuum source, \object{VLA~1}, located
 $\sim15\arcsec$ southwest of VLA~2. Later radio observations by
\citet{ang96} and \citet{gir97} led to discard VLA~1 as a  young stellar
object (YSO) associated with the CO multipolar outflow. As  discussed by
\citet{Gir09}, VLA~1 is not associated with the high-density molecular gas,
lying outside, near the border, of the NH$_3$ structure  encompassing the
radio continuum system VLA~2, and suggest that VLA~1 is a line-of-sight radio
source unrelated to the outflow.
However, the nature of the radio source VLA~1 has not been settled up to date. 
\citet{Car08} reanalyse multiepoch, centimetre and millimetre data of L723. 
Their data are compatible with VLA~1 belonging to the L723
cloud and being a radio-emitting, optically obscured T-Tauri star. 

Interestingly, we found in all our {\it H}- and {\it K}-band images a
stellar-like, near-infrared counterpart with coordinates
$\alpha=19^h~17^m~52\fs93$,
$\delta=+19^\circ~12\arcmin~8\farcs8$, which coincided within
$0\farcs05$ with the position of VLA~1 (\citealp{Car08}). 
This counterpart (see Fig.\ \ref{imagecomp1}) is also
detected in the four IRAC channels (3.6, 4.5, 5.8, and 8.0 $\mu$m)  of the
Spitzer images mentioned in Sect. 4.  
We performed aperture photometry of
the VLA~1 near-infrared counterpart for all the detected bands and found
magnitudes of 18.6, 17.0, 15.0, 14.9, 14.8, and 14.8 in the {\it Hc}, {\it
K$_c$}, 3.6, 4.5, 5.8, and 8.0 $\mu$m images,  respectively (errors in magnitude
are estimated to be $\leq 0.1$ mag). 
All magnitudes were calibrated relative to Vega (see \eg\ \citealp{rea05} for
IRAC). We combined these data to obtain near-infrared colours for the VLA~1
counterpart.
Following the current classification criteria (see \eg\ \citealp{fan09}, and
references therein), we found that the colours of this source are consistent
with those of an evolved  YSO of Class III. In particular, the position of this
source in the
[5.8]--[8.0] vs.\
[3.6]--[4.5] colour diagram corresponds to the loci where the Class III YSOs 
are located, and is also  compatible within the errors with the loci of the transition 
disc sources. 
In contrast, the VLA~1 counterpart lies outside the diagram region where the
background extragalactic sources are located. From these data we conclude that
VLA~1 is tracing the emission of an evolved pre-main sequence star of L723, as
proposed by \citet{Car08}. However, VLA~1  is likely unrelated to the molecular
outflows of L723, because the powering sources of these outflows need to be YSOs
in an earlier evolutionary  stage  than Class II/III.

\section{Summary and conclusions}

We present near-infrared images of the L723 field, obtained with  narrow-band
filters centred on the \fe\ $\lambda$ 1.644 $\mu$m and \hmol\  lines, together
with the off-line filters {\it H}$_c$ and {\it K}$_c$. The images cover an area
of  $4\farcm3 \times 4\farcm3$, thus include all the line-emission nebulae
associated with the HH~223 outflow that were detected  in H$\alpha$. The
analysis of the near-infrared images lead us to  the main results summarized
below.

\begin{enumerate} 

\item
We detected H$_2$ emission in a set of elongated nebular structures that appear
distributed from the southeast to the northwest of the L723 field, extending  
$\sim5\farcm5$ ($\sim0.5$ pc for a distance of  300~pc), which is reminiscent of a
parsec-scale outflow. The H$_2$ structures are found projected on the lobes of
the larger, east-west CO outflow, with an S-shape morphology. Several
H$_2$ structures are new identifications from this work. 

\item
Additional off-line filter images revealed that there is no significant
contribution from the continuum to the emission of  the nebular  
structures detected in the {\it K} band, except in HH~223-K1, the structure
closest to the position of the radio continuum system VLA~2.

\item 
Emission from the \fe\ 1.644 $\mu$m line was only detected in  HH~223 at the centre of the imaged field. Additional significant 
emission in the {\it H} band,  which mainly arises from continuum, was detected
in HH~223-K1.

\item 
We compared the appearance of the HH~223 outflow in the optical (H$\alpha$ and
\sii\ lines) and near-infrared (\fe\ at 1.644 $\mu$m, and H$_2$ at 2.122 $\mu$m 
at 4.5 $\mu$m) wavelength ranges, looking for the nature of the emissions.

\begin{itemize} 
\item
In general, the HH~223 outflow shows a similar large-scale morphology when the
emissions from molecular (at 2.122 and  4.5 $\mu$m) and atomic (H$\alpha$)
hydrogen lines are compared. Some of the discrepancies found among  the three
bands could be attributed to extinction effects  (\eg\ the non detection of
H$\alpha$ counterpart from the near-infrared nebulae close to the radio
continuum multiple system VLA2, and the larger extension of the 4.5~$\mu$m
emission compared with the emission at 2.122~$\mu$m, in HH~223-K1). Other
discrepancies  are  better explained from the physical conditions and kinematics
of the region  where they appear: \eg\ the non detection of an H$_2$ counterpart for
the H$\alpha$ emission at HH~223-B. 

\item  
The \sii\ and \fe\ emissions are well coincident for all the knots of 
HH~223  (the only structure of the outflow where near-infrared, 
line-emission from ionized gas was detected). The low-brightness emission
surrounding the knots is more extended in \sii\ than in \fe. We attribute this
to the lower velocity and density found around the knots, which is unable to excite
the \fe\ emission.

\item
Some differences are found when  the \fe\ and H$_2$~2.122$~\mu$m 
emissions are compared. The differences can be better attributed to the
different physical conditions required for these lines to emit (\eg\
velocity and density) than to be caused by extinction.

\end{itemize}

\item
We identified the near-infrared counterpart of the radio source VLA~1 in the
{\it H}, {\it K} and the four IRAC bands. The position of this source in the
colour-colour diagrams indicates that VLA~1 is tracing the emission of an
evolved (Class III) YSO. Thus, VLA~1 is likely unrelated to the H$_2$ outflow.

\item
Finally, we discussed the nature of the near-infrared emission for the
nebulae closest to the radio continuum sources, which is a candidate to power the
molecular and optical outflows. We propose that the H$_2$~2.122$~\mu$m emission
in HH~223-K1 and K2 is tracing shock-excited outflow gas, which also was
detected in SiO \citep{Gir09}, because the emission peaks of the  K1 and K2
near-infrared structures are coincident with two enhancements of the SiO
emission. In addition both H$_2$ and SiO emissions appear elongated
along a similar direction (PA~$\sim140^\circ$). 
The extended continuum emission associated with HH~223-K1  could be
tracing the walls cavity opened by the outflow.
\end{enumerate}

In summary, we propose that the H$_2$ nebular structures detected in the L723
dark cloud are most probably  associated with the radio continuum system
of protostellar sources found in this dark cloud.  All these H$_2$ nebular
structures could form part of a S-shaped outflow that also have an optical 
counterpart in the regions with low visual extinction. 

The knotty structures of the HH~223 outflow could be tracing internal working
surfaces of shocks that originate because the gas is  ejected at varying
speeds and/or with a varying direction and faster ejecta overtake slower ones. A
variable ejection velocity and an undulating jet morphology is expected when
the exciting outflow source belongs to a binary system.  The low-brightness, 
more diffuse line-emission structures could be tracing slow shocks, excited by
the interaction of the CO outflow with the accelerated gas of the walls cavity
opened by it.  The emission from the \fe\  line could be tracing the densest and
high-ionized regions in the outflow. 

The current data still do not allow us to discern which of  the radio continuum
component of the VLA~2 system is powering the large scale,
near-infrared/optical  HH~223 outflow. However at small scales (\ie\ short
dynamic time scales), the near-infrared emission and the SiO outflow seem to be
powered by SMA1. The large-scale scenario can be better settled by analysing the
kinematics and excitation conditions in the line-emission nebulae and needs to
be  derived from near-infrared spectroscopy.

\begin{acknowledgements} 
J.A.A.-P. is partially supported by the Spanish MICINN grant AYA2004-03136.
C.C.-G., R.E., and R.L. are supported by the Spanish MICINN grant
AYA2008-06189-C03.
C.C.-G. acknowledges support from MEC (Spain) FPU fellowship, FEDER funds, and
partial support from Junta de Andaluc\'{\i}a (Spain).
This publication makes use of data products from the Two Micron All Sky
Survey, which is a joint project of the University of Massachusetts and the
Infrared Processing and Analysis Center/California Institute of Technology,
funded by the National Aeronautics and Space Administration and the National
Science Foundation.\\
R.L. acknowledges the hospitality of the Instituto de Astrof\'{\i}sica de
Canarias, where part of this work was done.\\
\end{acknowledgements}

\end{document}